\documentstyle[12pt]{article}

\newcommand{\be}{\begin{equation}}
\newcommand{\ee}{\end{equation}}
\newcommand{\bea}{\begin{eqnarray}}
\newcommand{\eea}{\end{eqnarray}}
\newcommand{\lb}{\label}

\newcommand{\f}{\frac}

\normalbaselines
\begin{document}
\begin{flushright}
Freiburg THEP-95/2\\
quant-ph/9501004
\end{flushright}
\begin{center}
\vspace*{1.0cm}

{\bf IRREVERSIBILITY IN QUANTUM FIELD THEORY}\footnote{To appear in
{\em Nonlinear, dissipative, irreversible quantum systems},
edited by H.-D. Doebner, V. K. Dobrev, and P. Nattermann
(World Scientific, Singapore, 1995).}

\vskip 1.5cm

{\large {\bf Claus Kiefer}}

\vskip 0.5 cm

Fakult\"at f\"ur Physik \\
Universit\"at Freiburg \\
Hermann-Herder-Str. 3 \\
D-79104 Freiburg, Germany

\end{center}

\vspace{1 cm}

\begin{abstract}
It is shown how the programme of decoherence can be applied
in the context of quantum field theory. To illustrate the
role of gauge invariance, we first discuss
 the charge superselection rule in quantum
electrodynamics in some detail. We then present an example
where macroscopic electromagnetic fields are ``measured" through
interaction with charges and thereby rendered classical.
\end{abstract}

\vspace{1 cm}

A central role in our understanding of quantum theory as a physical
theory is played by the attempt to recover consistently from it
the classical appearance of our world. Assuming that quantum theory is
universally valid, a straightforward application of the superposition
principle leads to the occurrence of many superposed classical worlds
(i.e., many macroscopic different components of the total wave function),
in striking contrast to our everyday experience of {\em one} classical
world. This apparent contradiction has motivated von Neumann more than
sixty years ago to impose by hand a phenomenological law
{\em in addition to} the well-understood unitary time evolution
of quantum states -- the by now infamous ``collapse of the wave function".
Up to quite recently, this additional rule was indeed only applied as
an ad hoc prescription which works for all practical purposes but
which lacks any explanation in terms of some fundamental law.
Recently, however, there have been suggestions to ``put the
collapse into the equations, not just the talk" \cite{Pe}.
Typically, such explicit dynamical collapse models are of a stochastic
nature and lead to the irreversible emergence of ``events".

While most contributions to this conference can be, more or less,
adjoined to an approach of this kind, I shall here pursue a different
route which does not necessarily have to invoke some collapse mechanism and
 which has also aroused much interest in the past decade --
the attempt to understand the irreversible emergence of classical
behaviour through interaction with the environment (``decoherence"),
see, for example \cite{Zu,Ze93,FESt}, and the references therein.
It is the purpose of my contribution to report on some recent
applications of decoherence in a field-theoretic context
\cite{Ki92,GKZ}. This presents some novel features over and above
those already present in quantum mechanical systems, to which
most discussions have been restricted up to now. These novel features
are not only concerned with the much more sophisticated technical
nature of quantum field theories, but also with conceptual aspects
related to the presence of gauge invariance (and, in general
relativity, diffeomorphism invariance). After a brief introduction
into the general aspects of decoherence, I shall thus present
a discussion of the connection between symmetries and superselection
rules in quantum field theory and use as an illustration
the case of the
charge superselection rule in QED. I will then proceed to discuss
an example where macroscopic field strengths decohere through
their interaction with charges.

The basic observation for the understanding of decoherence is
provided by the fact that macroscopic systems cannot be considered,
not even approximately, as being isolated from their natural
environment \cite{Ze70}. In fact, they are strongly quantum
correlated with it. Traditional discussions of the measurement
process consider a quantum mechanical system, ${\cal S}$
(described by a basis of states $\{\varphi_n\}$), coupled to an
``apparatus", ${\cal A}$ (described by a basis $\{\Phi_k\}$).
In the well-known example by Hepp \cite{He}, ${\cal A}$ consists of an
infinite chain of spin 1/2 particles. A measurement is there
considered as complete only in an (unphysical) limit of infinite
time, and only with respect to an a priori choice of local
observables, see the criticism in \cite{Be} and \cite{La}.

Taking now into account the natural environment, ${\cal E}$
(described by a basis $\{ {\cal E}_l\}$), of the
apparatus, phase relations between different states of the apparatus
become delocalised through correlations with the {\em huge}
number of environmental degrees of freedom (photons, air molecules,
\ldots). Tracing them out in the total, quantum-entangled, state
(I consider the simplest case of a correlation)
\be \vert\Psi\rangle =\sum_n c_n\vert\varphi_n\rangle \otimes
 \vert\Phi_n\rangle\otimes \vert{\cal E}_n\rangle \lb{tot} \ee
leads to a reduced density matrix for ${\cal A}$ of the form
\bea \rho_{\cal A}& = & \mbox{Trace}_{\cal E}
       \vert\Psi\rangle\langle\Psi\vert    \nonumber\\
  & = & \sum_{n,m} c_n^*c_m \vert\varphi_m\rangle\otimes
  \vert\Phi_m\rangle\langle{\cal E}_n\vert{\cal E}_m\rangle
  \langle\varphi_n\vert\otimes\langle\Phi_n\vert \nonumber\\
  & \approx & \sum_n \vert c_n\vert^2 \vert\varphi_n\rangle
  \otimes\vert\Phi_n\rangle\langle\Phi_n\vert\otimes \langle
  \varphi_n\vert, \lb{rho} \eea
where the last step follows from the approximate orthogonality
of different environmental states (which is what happens in realistic
cases). Thus, the density matrix (\ref{rho}) assumes the form of
an {\em approximate} ensemble, and it seems {\em as if} the system
has ``collapsed" into one of the states $\varphi_n$ with a
probability $\vert c_n\vert^2$.

If initially there is no (or almost no) quantum entanglement between
${\cal A}$ and ${\cal E}$, the local entropy
\be S= -k_B\mbox{Tr}(\rho_{\cal A}\ln \rho_{\cal A}) \lb{en} \ee
will increase by this interaction -- classical properties
emerge in a practically {\em irreversible} manner, since in realistic cases
the environmental degrees of freedom never return to their initial
state because of the enormous Poincar\'{e} times usually
involved.

I must emphasise that the result (\ref{rho}) does not yet imply the
observation of a definite measurement outcome -- only the
interference terms have locally disappeared, and the total state (\ref{tot})
is still a pure state. To explain the occurrence of {\em one}
measurement result, one must adhere to one of the following
options. The first possibility is that the total state is {\em really}
 given by (\ref{tot}). In the framework of this ``many-worlds
 interpretation" facts emerge only through the locality
 of observers who have only a very restricted algebra of observables
 at hand. The second possibility has to invoke an explicit collapse
 mechanism for the total state in the sense mentioned at the
 beginning \cite{Pe}. A decision between these two options cannot yet be
 made and is to a large extent a matter of taste \cite{ZeB}.
It is, however, important to keep in mind that it is in principle
possible to distinguish between these options, since recoherence
would only be possible in the first case. In fact, it has drastic
consequences for the arrow of time in a recollapsing quantum universe
\cite{KZ}.

Let me now turn to QED and the charge superselection rule
\cite{GKZ}. This may also serve as a prototype for other gauge theories,
which are not discussed here.

Consider first the classical theory. Infinitesimal gauge
transformations parametrised by an arbitrary function
$\xi({\bf x})$ are generated by
\be Q^{\xi}=
    \int d^3x (E^a\partial_a\xi +\rho\xi), \lb{qxi} \ee
where $E^a$ denotes the components of the electric field strength,
 and $\rho$ is the charge density.
Integration by parts yields
\be Q^{\xi}=
    \int_{S_{\infty}}d\sigma n_aE^a\xi -\int d^3x\ \xi
     (\partial_aE^a-\rho). \lb{par} \ee
The surface integral is over $S_{\infty}$, the ``sphere at infinity",
and $n_a$ is the outward pointing normal.
An important feature in electrodynamics, which is connected
with the presence of gauge symmetry, is the Gauss constraint
equation,
\be {\cal G}\equiv \partial_aE^a -\rho =0. \lb{gau} \ee
Consequently, on the constraint surface,
\be Q^{\xi}\vert_{{\cal G}=0} =\xi\int_{S_{\infty}}
    d\sigma n_aE^a \equiv \xi\ Q, \lb{cha} \ee
where $Q$ denotes the total charge. It is an
observable in the formal sense, since it commutes with ${\cal G}$ on the
constraint surface.

 In the quantum theory the above relations remain
formally valid as operator equations. The charge $\rho$ is then
given by $-ie\pi_{\psi}\psi$, where $\psi$ is the spinor field,
and $\pi_{\psi}$ its conjugate momentum.
If quantisation is performed in the functional Schr\"odinger picture
\cite{Ja}, the constraint (\ref{gau}) is implemented as a restriction
on physically allowed wave functionals, $\Psi[A_a,\psi]$, as
\be \partial_a\frac{1}{i}\frac{\delta\Psi}{\delta A_a}
     =-ie\psi\frac{\delta\Psi}{\delta\psi}. \lb{cons} \ee
This equation expresses the simultaneous invariance of the wave functional
with respect to local gauge transformations of the vector potential
and the spinor field.

The Gauss constraint only generates {\em asymptotically trivial}
gauge transformations, as can be seen from (\ref{par}).
How should one interpret the remaining gauge transformations?
This poses the question on the physical meaning of $S_{\infty}$
and, thus, the role of infinity in this discussion.
One can distinguish between two possibilities. First,
$S_{\infty}$ may lie outside a quantum mechanically closed universe,
which means that (although space itself may be finite or infinite)
there are no degrees of freedom outside the sphere. In this case
$Q^{\xi}$ should generate redundancy transformations and thus
only allow an eigenvalue $Q=0$ of the total charge operator.
In cosmology, this would be a sensible result!

In the second possibility, $S_{\infty}$ lies far away for all
practical purposes, but there may still be charges and/or fields
outside. In this case $S_{\infty}$ may serve as a reference
system (compare \cite{AS}), and $Q^{\xi}$ should generate
meaningful {\em symmetries}. The total state may be in a charge
eigenstate or not. If it is, for example, in a superposition
of two states with negative and positive elementary charges,
respectively, $\Psi\equiv\Psi_++\Psi_-$, the action of the
charge operator would be as follows,
\be e^{i\hat{Q}}\Psi= e^{ie\xi_{\infty}}\Psi_+
    +e^{-ie\xi_{\infty}}\psi_-, \lb{sup} \ee
where $\xi_{\infty}$ is the value of the function $\xi({\bf x})$
at ``infinity". An example of an observable (i.e., of
a self-adjoint operator which commutes with the Gauss
constraint and thus is invariant under local gauge transformations)
which has non-vanishing matrix-elements between both charge ``sectors",
is the Mandelstam observable\\ $\exp(ie\int_{-\infty}^{\bf x}{\bf A}d
{\bf s})$. Any (quasi-) local observable, however, {\em commutes}
with the total charge, cf. (\ref{cha}), where only $S_{\infty}$
is involved, since it only has support {\em inside}
the ``sphere at infinity". This fact is often referred to as the
{\em charge superselection rule} \cite{SW}. Locally, the state
is {\em indistinguishable} from a mixture of states, although
the total state may be pure (see \cite{HLM} for a recent discussion
of this in the framework of consistent histories).
Due to Gauss law, charges have always been ``measured" by the asymptotic
fields and are thus always ``decohered" with respect to
bounded subsystems.

In algebraic field theory one often considers a special case of
the second of the above options, that fields may be outside the
sphere $S_{\infty}$, but no sources. For such an ``island universe"
one can consistently restrict attention to one decohered component
of a superposition like (\ref{sup}).

Thus, it is physically more relevant to discuss local
superpositions of charges, independent of whether the total state
of the Universe is in a charge eigenstate or not \cite{GKZ}.
An interesting question is, for example, over what distances an
electronic wave packet can be split and {\em coherently}
re-unified. Since the influence of the Coulomb field acts
in a reversible manner, the answer depends on the strength of the
{\em irreversible} interaction with the radiation field.
Are there quantitative estimates? Joos and Zeh \cite{JZ}
have demonstrated that thermal radiation affects free electrons
very efficiently by Thomson scattering. For example, if an initially
separating state between electron and field is assumed, there
remains (for a temperature $T=300 K$ of the electromagentic field)
 after one second a coherence length for the
electrons of only $0.1 cm$ (the dependence of the coherence length
on time is as $t^{-1/2}$). Even more effective seems to be the
influence of the electron's own radiation field, although
there is not yet a definite conclusion about the quantitative outcome
\cite{GKZ}.

At this point I would only like to mention analogous examples in
other theories, such as the mass superselection rule in general
relativity, which can be understood along the lines presented here
\cite{GKZ}.

Due to the mutual interaction of charges and fields, one can not
only discuss the measurement of charges by fields, but also the
opposite case of a field measurement by charges. It depends of course
on the experimental situation, which aspect is the more important one.
In fact, a detailed investigation of the field measurement by
charges was crucial in the seminal work of Bohr and Rosenfeld.

One may wish to consider, for example, a macroscopic superposition
of two electric fields, one pointing upwards, and the other pointing
downwards. The total state may be written in the semiclassical
form \cite{Ki92}
\be \Psi[\psi\psi^{\dagger},A]\approx
  e^{-iVAE}\chi +e^{iVAE}\chi^*, \lb{supf} \ee
where $V$ is the space volume, $E$ and $A$ are the respective components
of the electric field and the vector potential, and $\chi$ is the
state of the electrons. The whole discussion is performed within
the functional Schr\"odinger picture of QED \cite{Ja}. In the simplest case,
$\chi$ is assumed to be in a Gaussian state (corresponding to
a generalised, $A$-dependent, vacuum state), but it is
straightforward to consider more complicated states. The reduced
density matrix for the electric field can be obtained from
(\ref{supf}) by tracing out the degrees of freedom corresponding
to the electrons, cp. the general expression (\ref{rho}).
One finds for the non-diagonal elements of the reduced
density matrix (apart from phase factors)
\be \rho_{\pm} =e^{2iVAE}\mbox{Trace}_{\psi,\psi^{\dagger}}\chi^2
 \approx e^{2iVAE}\exp\left(-\f{Ve^2E^2}{512\pi m}\right), \lb{rhopm} \ee
in which the limit $t\gg m/eE$ (which is rapidly
reached) was performed. Note that the interaction with the
charge states leads to an exponential suppression factor of
the corresponding interference terms for the field; in the infrared
limit of $V\to\infty$ one finds exact decoherence. In realistic
cases, however, a finite coherence width remains, so one can
in principle subject these results to experimental confirmation.
For an electric field of $E\approx 10^7$ Volts per centimetre,
for example, one finds that interference effects are observable
on length scales $L\leq 10^{-4}$ centimetres.

Thus, in summary, the programme of decoherence can successfully
be applied in the context of quantum field theory, and one can understand
the irreversible emergence of classical properties for
quantities such as electric charge, mass, or macroscopic
field strengths. Of course, a necessary input is the assumption
of special initial states of low entropy
(i.e., the absence of initial correlations), such that the
local entropy (\ref{en}) for relevant subsystems can increase.
This leads eventually into the realm of cosmology and the
subject of quantum gravity \cite{KZ,ZeT}.

\section*{Acknowledgments}
I owe my thanks to Domenico Giulini and H.-Dieter Zeh for
collaboration and many invaluable discussions.

\end{document}